Revisiting the Carrington Event: Updated modeling of atmospheric effects

Brian C. Thomas, Keith R. Arkenberg, and Brock R. Snyder II

Washburn University
Department of Physics and Astronomy
1700 SW College Ave.
Topeka, KS 66621

Abstract:

The terrestrial effects of major solar events such as the Carrington white-light flare and subsequent geomagnetic storm of August-September 1859 are of considerable interest, especially in light of recent predictions that such extreme events will be more likely over the coming decades. Here we present results of modeling the atmospheric effects, especially production of odd nitrogen compounds and subsequent depletion of ozone, by solar protons associated with the Carrington event. This study combines approaches from two previous studies of the atmospheric effect of this event. We investigate changes in $NO_y$ compounds as well as depletion of $O_3$ using a two-dimensional atmospheric chemistry and dynamics model. Atmospheric ionization is computed using a range-energy relation with four different proxy proton spectra associated with more recent well-known solar proton events. We find that changes in atmospheric constituents are in reasonable agreement with previous studies, but effects of the four proxy spectra used vary more widely than found by one of those studies. In particular, we find greater impact for harder proton spectra, given a constant total fluence. We report computed nitrate deposition values and compare to measured values in ice cores. Finally, we briefly investigate the impact of the modeled ozone depletion on surface-level solar ultraviolet radiation.

1. Introduction:

Effects on the Earth due to rare but very energetic solar proton events (SPEs) are of interest for a variety of reasons. The solar flare of August-September 1859, commonly know as the Carrington event, is the most intense white light flare observed in the past 450 years [*Tsurutani et al.*, 2003; *Clauer and Siscoe*, 2006; *McCracken et al.*, 2001]. It is presumed that a prior solar event, probably on 27 August 1859, launched a coronal mass ejection that resulted in geomagnetic storms on 28 and 29 August [*Smart et al.*, 2006]. The storm was so intense that induced currents in telegraph lines arced, sparking fires in both the United States and Europe [*Loomis et al.*, 1861], and producing visible aurorae as far south as Cuba and Jamaica [*Tsurutani et al.*, 2003]. According to *Smart et al.* [2006], the SPE was probably of a class of shock-dominated events consisting of two major intensity peaks; the estimated duration of the event is thirteen days, concluding on 9 September 1859.

Intense SPEs like that associated with the Carrington event are increasingly of interest, especially with the likelihood that the Sun is leaving a Grand Solar Maximum. *Barnard et al.* [2011] predict that while there will be fewer flares and coronal mass ejections, those that do happen will be more intense and of higher fluence. *Owens et al.* [2011] estimate that the probability of a large event during the next solar cycle (number 24) is 0.8; this compares to a probability of 0.5 for the previous cycle. Therefore, another event similar to the Carrington appears likely in the coming decades.

The Carrington event has received significant attention in recent years, including an entire special issue of the journal *Advances in Space Research* [*Clauer and Siscoe*, 2006]. Those authors and others have considered various effects, including the impact on Earth-orbiting satellites [*Odenwald et al.*, 2006]; ionizing radiation doses for crews of manned missions [*Townsend et al.*, 2003]; and effects on HF communication due to changes in the ionosphere [*Rodger et al.*, 2008]. Another effect of interest is changes in atmospheric chemistry, especially the production of odd-nitrogen oxides (denoted $NO_y$) by SPE-induced ionization in the middle atmosphere, followed by depletion of stratospheric ozone. *Thomas et al.* [2007] used a two-dimensional atmospheric model, with ionization rate profiles taken from the October 1989 SPE scaled up to an estimated fluence for the Carrington event. *Rodger et al.* [2008] used a one-dimensional model for a similar study, but instead computed atmospheric ionization rate profiles using several different proxy spectra and an estimated intensity-time profile from *Smart et al.* [2006].

In this work, we combine the approaches of those two studies, using a two-dimensional model with ionization rate profiles similarly (and independently) computed from proxy spectra and the *Smart et al.* [2006] intensity-time profile, which is based on ice core nitrate levels analyzed by *McCracken et al.* [2001]. Primarily, we wish to refine and improve understanding of the atmospheric effects, particularly changes in ozone, due to the Carrington event. Both previous studies have limitations due to the different approaches to computing ionization rate profiles and the different atmospheric models used. While *Rodger et al.* [2008] used what should be a more accurate approach to the ionization rate profiles, their one-dimensional model has limitations in both altitude range, and, of course, in the fact that it is restricted to one geographic location. On the other hand, while *Thomas et al.* [2007] used a simple scale factor to estimate the ionization rate profile, the two-dimensional model used extends from the ground to 116 km and covers the full latitude range, which is necessary to evaluate the large-scale and long-term effects of the event.

In addition to studying the atmospheric effects, particularly on ozone, we have also computed values for nitrate deposition in ice cores, which can be compared to existing and future measured values. Finally, we have briefly investigated the enhancement of solar-ultraviolet (UV) radiation at the Earth's surface due to the reduction in column $O_3$.

2. Modeling the flux and intensity-time profile of the Carrington event
Several possibilities exist for modeling the flux and intensity-time profile of the Carrington SPE. *Smart et al.* [2006] argue that spectra of the August 1972 and March 1991 are likely to be representative of the Carrington event. These are relatively soft spectra. *Rodger et al.* [2008] used these spectra, as well as two harder spectra – those of the September and October 1989 SPEs. We will follow suit and model the effects of all four proton spectra.

Following *Rodger et al.* [2008] we use a Weibull distribution fit for the SPE spectra. The differential flux can be expressed as

$$\frac{d\Phi(E)}{dE} = \Phi_0 k\alpha E^{\alpha-1} e^{-kE^{\alpha}},$$

where $E$ is the proton energy in MeV, and $\Phi_0$, $k$, and $\alpha$ are fitting parameters. The differential flux values are expressed in units of protons cm$^{-2}$ s$^{-1}$ sr$^{-1}$ MeV$^{-1}$. Values for $k$ and $\alpha$ are taken from *Xapsos et al.* [2000] and are listed in Table 1. The four spectra are shown in Figure 1 as normalized differential fluence as functions of energy. In our study, $\Phi_0$ is taken as a function of time, introducing the intensity-time profile based on that of *Smart et al.* [2006] (with specific values kindly provided by Craig Rodger [*personal communication*]). This intensity-time profile is shown in Figure 2. Note that the same profile is used for all modeled spectra; in each case we normalize to a total $E > 30$ MeV proton fluence of $1.9 \times 10^{10}$ cm$^{-2}$ [*McCracken et al.*, 2001; *Rodger et al.*, 2008].

3. Calculation of atmospheric ionization rate profiles

We compute the rate of ionization in the atmosphere as a function of altitude and time, following the approach of *Jackman et al.* [1980] (see also *Verronen et al.* [2005]). A separate code is used for this computation and the results are then fed to the atmospheric model. Here we describe the details of this calculation, following the Appendix of *Jackman et al.* [1980].

The atmosphere from 0-116 km is divided into bins of approximately 2 km thickness. The energy range 1-10$^5$ MeV is divided into 60 evenly spaced logarithmic bins. We assume an isotropic distribution of incident particles; the pitch angle range from 0-$\pi$/2 radians is divided into 35 equal bins.

The energy deposited in altitude bin $i$ by protons with kinetic energy $E$ and pitch angle $\theta$ is given by

$$E_{di}(\theta, E) = E - \left\{-\frac{\Delta z_i}{A}\sec\theta + E^B\right\}^{1/B} \text{MeV},$$

where we have used the range energy relation [*Bethe and Ashkin*, 1953; *Whaling*, 1958; *Sternheimer*, 1959; *Green and Peterson*, 1968]:

$$R(E) = A \left(\frac{E}{1 \text{ MeV}}\right)^B \text{ gm cm}^{-2}$$

with $A = 2.71 \times 10^{-3}$ and $B = 1.72$ for $1 \leq E \leq 1550$ MeV, and with $A = 0.834$ and $B = 0.94$ for $E > 1550$ MeV. These fits are good to within 16% at energies up to $10^5$ MeV and to within 5% at most energies [*Jackman et al.*, 1980].

The energy deposited by an isotropic flux of monoenergetic protons over the upper hemisphere is then found by integrating

$$E_{di}(E) = \int_0^{2\pi} \int_0^{\pi/2} \cos\theta \, E_{di}(\theta, E) \sin\theta \, d\theta d\phi \quad \text{MeV sr}.$$

Finally, using the differential flux discussed above, the total energy deposited in altitude bin $i$ is

$$E_{di} = \int_{E_{min}}^{E_{max}} \frac{d\Phi(E)}{dE} E_{di}(E) \, dE \quad \text{MeV cm}^{-2} \text{ s}^{-1}.$$

For this calculation we use $E_{min} = 1.0$ MeV and $E_{max} = 10^5$ MeV. The ionization rate at altitude bin $i$ is then just $E_{di}/(35 \text{ eV})$, where 35 eV is the energy needed to produce one ion pair [*Porter et al.*, 1976]. This calculation results in ionization rate as a function of altitude and time, since here we use a time-dependent proton flux. In Figure 3 we show the volume ionization rate for each SPE case considered. The calculations described above result in ionization rate per area; for this figure we have converted those values to ionization rate per volume, assuming altitude bins of 2 km thickness (height). For simulation runs this conversion is done using time- and location-dependent altitude bin thickness computed by the atmospheric model.

4. Atmospheric modeling

We use the Goddard Space Flight Center (GSFC) two-dimensional atmospheric chemistry and dynamics model, which has been used in several previous studies of SPEs [*Jackman et al.*, 1990, 1995, 2000, 2001, 2005a, 2005b; *Thomas et al.*, 2007]. The model has also been used for other high-energy events such as supernovae [*Gehrels et al.*, 2003; *Thomas et al.*, 2008] and gamma-ray bursts [*Thomas et al.*, 2005; *Ejzak et al.*, 2007]. In addition, it has been used to study the atmospheric effects of enhanced high-energy cosmic ray flux [*Melott et al.*, 2010]. Here we briefly describe the model; more details can be found in *Thomas et al.* [2005] and references therein. The model's two spatial dimensions are altitude and latitude. The latitude range is divided into 18 ten-degree bands, from pole to pole. The altitude range extends from the ground to approximately 116 km, with 58 evenly spaced logarithmic pressure levels. The model computes 65 constituents with photochemical reactions, solar radiation variations, and transport, including winds and small scale mixing. The version of the model that we use here has a time step of one day; daily averaged constituent values are computed. We perform all model runs with pre-industrial values for constituents such as CFCs, so as to focus on the effects of solar protons rather than any anthropogenic changes.

The ionization rate profiles, computed as described above, are input to the atmospheric model as a function of altitude and time. The duration given by the *Smart et al.* [2006] intensity-time profile is 13 days. In *Thomas et al.* [2007] it was assumed that the solar protons would be restricted to latitudes > 60° and ionization was input uniformly over these latitude bins in the atmospheric model. In this study we have computed a latitude-dependent scale factor using the approach of *Elasser et al.* [1956]. The scale factor at a given latitude $\theta$ and longitude $\phi$ is given by:

$$f = 1 - \cos^4(\lambda_G)$$

where $\lambda_G$ is the geomagnetic latitude, given by:

$$\sin \lambda_G = \sin(\theta_P)\sin(\theta) + \cos(\theta_P)\cos(\theta)\cos(\phi_P - \phi)$$

where $\theta_P$ and $\phi_P$ are the latitude and longitude of the North geomagnetic pole. According to *Shea and Smart* [2006], the location of the North geomagnetic pole in 1850 was 78.62° North, 296.40° East. The scale factor $f$ was computed using these values for 360 points in longitude, and 18 points in latitude (from 85° North to 85° South, corresponding to the latitude bins of the atmospheric model), then averaged over longitude and used to scale the ionization rate profiles for use in the atmospheric model. The resulting scale factor as a function of latitude is shown in Figure 4.

The ionization rate profiles are treated as a source of odd-hydrogen oxides, $HO_x$ (e.g., H, OH, $HO_2$), and odd-nitrogen oxides, $NO_y$ (e.g., N, NO, $NO_2$, $NO_3$). These compounds deplete ozone through several catalytic cycles [*Jackman and McPeters*, 2004]. It is assumed that 1.25 $NO_y$ molecules are produced for each ion pair at all altitudes [*Porter et al.*, 1976]. We use a production rate of $HO_x$ as a function of altitude from *Solomon et al.* [1981], with maximum production of 2.0 molecules per ion pair. The production is most efficient between about 40 and 70 km. This, combined with the relatively short lifetime (hours) of $HO_x$ compounds in this part of the atmosphere [*Jackman and McPeters*, 2004], means that the impact of these compounds on total column ozone is small. $NO_y$ compounds, on the other hand, are produced in relatively large amounts in the stratosphere, and those produced above 50 km can be transported downward. These compounds have lifetimes of months in the upper stratosphere and years in the lower stratosphere, and can therefore have a significant impact on total column ozone [*Jackman et al.*, 2005b].

5. Results

Here we present results of our modeling of the four proxy cases described above. For each case the model was run for 10 years. Ionization was input starting at 27 August. Our primary results are changes in $NO_y$ and $O_3$. Most values discussed here are given in terms of point-wise percent difference between the perturbed run and a background run without ionization input.

5.1 Changes in $NO_y$ and $O_3$

Figure 5 shows the percent difference in column $NO_y$ between perturbed and unperturbed runs for each proxy case over the first 20 days and Figure 7 shows the same data over the first 300 days, where time zero corresponds to 27 August 1859. Similarly, Figure 6 and Figure 8 show the percent difference in column $O_3$. Some features are worth noting. First, the greatest change in $NO_y$ coincides with the highest ionization intensity (see Figure 3) and then decreases over time. Second, the changes follow the latitude distribution given by the scale factor (greatest intensity in the polar regions; see Figure 4). Note, however, that the effects are not exactly symmetric in latitude. The $NO_y$ increase is greatest in the Southern Hemisphere, for both the short and long term. This is due to a relative seasonal difference between the hemispheres; while it is late summer in the Northern Hemisphere, it is late winter in the Southern Hemisphere. The presence or absence of sunlight at a given location and time has a major effect on the atmospheric conditions since photolytic reactions affect the relative abundance of several constituents of the $NO_y$ family. The absence of sunlight prevents photolytic destruction of certain species, allowing the increase to be larger in the Southern Hemisphere in this case. For more details, see the discussion in Section 3.1 of *Thomas et al.* [2005].

Similar patterns are seen in changes in column $O_3$ (Figure 6 and Figure 8). However, it may be noted that the maximum decrease in $O_3$ actually occurs in the Northern Hemisphere in each case. This again is due in part to seasonal effects. In the winter, concentrations of NO and $NO_2$ in particular are elevated relative to their summer-time values. These compounds are the primary $NO_y$ species responsible for depletion of $O_3$, and hence that depletion is more effective in local winter. This effect is enhanced here by transport of $NO_y$ compounds produced at lower latitudes into the Polar Regions, which replenishes some of the NO and $NO_2$ that was previously destroyed by photolysis.

In Figure 9 we show an altitude profile of the change in $NO_y$, from the ground to the top of the stratosphere. While greater changes occur at higher altitudes, we are primarily interested in the impact on $O_3$, which is normally concentrated in the stratosphere and is therefore more affected by $NO_y$ that is produced in or transported to this part of the atmosphere. Figure 10 shows the corresponding change in $O_3$. Both figures are at day 100 after the start of the event, which is in mid-November. This corresponds roughly to the Northern Hemisphere maximum change in $O_3$ column density as shown in Figure 8. Accordingly, we see greater effects in the Northern latitudes for both $NO_y$ and $O_3$. An interesting feature is notable in Figure 10. While total column $O_3$ shows a decrease everywhere (Figure 8), the profile $O_3$ change is actually increased (relative to the unperturbed run) in the lower stratosphere. This is an effect known as "self-healing" [*Jackman and McPeters,* 1985]. Normally, ozone is concentrated between 30 and 40 km altitude and is sparse at lower altitudes. However, when $O_3$ is depleted in this range, solar UV (normally absorbed by the ozone) can penetrate to lower than normal altitudes. This UV then creates $O_3$ (by photolysis of $O_2$) at altitudes where normally there is little or none. Since these plots show a comparison between a run without ionization input (where little ozone exists at lower altitudes) and one with that ionization (where UV penetrates more deeply and creates lower altitude ozone), we see an increase at these lower altitudes. This effect partially mitigates the total column depletion. Note in Figure 10 that this "self-healing" effect is most pronounced in the Southern Hemisphere (especially around 40-50°) at this time. This corresponds to the fact that at this point in the simulation (mid-November) it is late spring in this hemisphere.

5.2 Differences in Proxy Case Results

We now consider differences in the atmospheric impact of the four proxy cases modeled here; this discussion will focus on features seen in the column density changes of $NO_y$ and $O_3$. Table 2 lists maximum values of several quantities for the different cases. (Note that each value listed corresponds to a specific set of location and timestep values.) Figure 11 shows the globally averaged percent difference in $O_3$ column density between perturbed and unperturbed runs for each proxy case over the first 1500 days. This may be thought of as an overall view of the intensity of the $O_3$ impact of each case. As is readily seen here, the two softer-spectrum cases (March 1991 and August 1972) have similar (and smaller) total impact on global $O_3$ column density. The harder-spectrum cases (October 1989 and September 1989) show a larger impact, with the September case significantly more effective than the October case. All cases return to the unperturbed atmospheric condition after about four years.

A few other interesting features appear. First, the maximum ionization rate is similar for all cases, but largest for March 1991. Similarly, the maximum increase in $NO_y$ column density (at a particular latitude and time, see Figure 5) occurs for this case. Conversely, March 1991 gives the smallest change in $O_3$ column density, both locally and globally averaged. A similar comparison can be made between October 1989 and September 1989. We can gain some insight into these patterns by looking at the normalized differential spectra presented in Figure 1, as well as the ionization rate profiles in Figure 3. The March 1991 case is relatively high in lower energy protons, compared to all the other cases, while it has fewer mid-range protons compared especially to August 1972, but is a middle case for higher energies. The ionization produced in this case is locally more intense than any of the other cases, and peaks around 75 km altitude (see Figure 3), ranging higher in altitude than the August 1972 case, and not as low in altitude as the 1989 cases. Overall, while the $NO_y$ change is greatest for March 1991, the $O_3$ change is relatively small because the peak ionization rate (and $NO_y$ production) is well above the stratosphere (this can also be seen in Figure 10). For comparison, the ionization in the August 1972 case, while locally less intense, peaks at an altitude of about 50 km, just at the top of the stratosphere. For the 1989 cases, the relatively large number of high energy protons leads to deeper penetration in altitude, which means less intense local ionization, but also a greater impact on $O_3$ since there is significant ionization produced throughout the stratosphere.

5.3 Comparison with Previous Studies

Since we intend this study to be a combination of methods used in *Thomas et al.* [2007] and *Rodger et al.* [2008], it is worthwhile to compare the present results to results in those papers. One nearly direct comparison that can be made is that of ionization rates for the different proxy cases. We find very good agreement between our results presented in Figure 3 and those given in Figure 4 of *Rodger et al.* [2008]. We consider this a validation of our ionization rate calculations, which were performed independently, but based on a similar theoretical approach.

In terms of atmospheric chemistry changes, we find qualitatively similar results to those in *Thomas et al.* [2007]. However, the two soft-spectrum cases here yield *smaller* globally averaged $O_3$ depletion than in that study, while the two hard-spectrum cases yield larger values. A similar comparison can be made for the maximum local depletion in column $O_3$. Interestingly, it appears that the depletion found in the previous study is a good "average" value between the proxy cases considered here. However, it must be noted that the fluence used here and in *Rodger et al.* [2008] ($1.9 \times 10^{10}$ cm$^{-2}$) is smaller than that used in *Thomas et al.* [2007] ($2.74 \times 10^{10}$ cm$^{-2}$). It may be expected, then, that $O_3$ depletion would be smaller here than that found in *Thomas et al.* [2007]. If *Smart et al.* [2006] are correct that spectra of the August 1972 and March 1991 are likely to be representative of the Carrington event, then the results we find for those cases here with a lower fluence appear to be consistent with the previous results using a higher fluence.

It is particularly interesting to note that the October 1989 values found here are quite similar to those found in the previous study [*Thomas et al.*, 2007], wherein ionization rates for this same SPE event were used as a proxy through a simple scale factor derived from ice core estimates of total proton fluence. This result suggests that such a scale-factor approach may be sufficiently accurate for this kind of study, allowing one to possibly bypass more complete calculations. We note, however, that this is based on only one data point; further studies could in principle check the generality of this result.

*Rodger et al.* [2008] concluded based on their modeling that the atmospheric response of the Carrington SPE is not strongly dependent on the event spectrum. In this study, however, we find a greater divergence between the four proxy cases, in particular between the soft and hard spectra, but also between the two different hard spectra (October and September 1989). The greatest difference appears between the March 1991 and September 1989 cases, where the September case gives 1.7 times the maximum globally averaged $O_3$ depletion and 1.5 times the maximum local $O_3$ depletion.

More detailed comparison with the results of *Rodger et al.* [2008] is possible, but limited. Their 1-dimensional model does not allow for comparison of latitude-dependent features (although they do present results for a location in the Northern Hemisphere as well as the Southern Hemisphere). Similarly, while their model extends to higher altitudes, it does not extend all the way to the ground, as our model does. This makes comparisons of overall $O_3$ impacts more difficult.

We can make a limited comparison of $O_3$ column density changes. In their Figure 9, *Rodger et al.* [2008] present percent change in total column $O_3$ above 30 km altitude, at both 70° North and South. It is important to note that 30 km is roughly the middle of the stratosphere, and so column changes above this altitude may be quite different than column changes that include the full altitude range. (See Figure 10 and discussion above). The maximum depletion they find ranges from about 7% to about 9.5%, depending on the proxy case. For both the North and South locations they find that March 1991 gives the smallest change. August 1972 and October 1989 produce similar changes, with the October case being somewhat worse in the North and the August case being nearly the same but slightly worse in the South. (They did not present results for September 1989.)

In Figure 12 and Figure 13 we present results that are the closest comparison we can make with Figure 9 of *Rodger et al.* [2008]. Here, we show percent change in total column $O_3$ (from 116 km to the ground) at latitude bands centered on 75° North and 75° South, respectively, over the first 20 days, where, as before, 0 corresponds to 27 August. The maximum percent change in $O_3$ during this time interval is similar to that in *Rodger et al.* [2008] (between about 5.5% and 9%), but we find greater difference between the four proxy cases, with March 1991 and August 1972 being quite similar and consistently smaller than the 1989 cases. That is, while *Rodger et al.* [2008] found the values for August 1972 and October 1989 to be similar here, in our results they are rather widely separated. This very likely due to the difference in altitude range between our models, since both of the 1989 cases produce significant ionization (and corresponding $O_3$ depletion) below 30 km. Another major difference is that in our results the $O_3$ depletion continues past the 20-day mark, whereas in *Rodger et al.* [2008] the $O_3$ column density begins to recover almost immediately (within a few days). This is most likely due to transport, which allows more of the $NO_y$ compounds to interact with and deplete $O_3$. The differences here highlight the need for modeling that includes the full altitude range (especially lower altitudes) as well as transport in looking at long-term, global effects of solar proton events.

5.4 Nitrate Deposition

The proton fluence assumed for the Carrington event has been derived from ice cores [*McCracken et al.* 2001]. The GSFC atmospheric model used here can compute deposition of nitrate by rain and snow. In *Thomas et al.* [2007] it was found that the deposition summed over three months following the event over the 10° latitude band centered at 75° North gave a value of 1360 ng cm$^{-2}$. It was also reported there that using Equation 2 in *McCracken et al.* [2001], with conversion factor $K = 30$, and proton fluence $2.74 \times 10^{10}$ cm$^{-2}$, gave an expected deposition of 822 ng cm$^{-2}$. It should be noted, however, that there is fairly wide uncertainty in the value of $K$, part of which may be due to spectral hardness [*McCracken et al.* 2001]. For $K = 40$ and fluence $2.74 \times 10^{10}$ cm$^{-2}$ the result is 1096 ng cm$^{-2}$.

In the present study we use a fluence of $1.9 \times 10^{10}$ cm$^{-2}$, which for $K = 30$ gives 570 ng cm$^{-2}$; for $K = 40$ the result is 760 ng cm$^{-2}$. Computing the total deposition for the current simulations over the same time period and spatial region gives 1091 ng cm$^{-2}$ for the September 1989 case, 1090 ng cm$^{-2}$ for the October 1989 case, and 1087 ng cm$^{-2}$ for both the March 1991 and August 1972 cases. Therefore, we find a similar overestimate in this study compared to *Thomas et al.* [2007]. Given all the uncertainties involved this discrepancy does not seem unreasonable; it may indicate that the model systematically over-estimates deposition, or that the values of $K$ are systematically too small. Finally, it is interesting to note that the softer spectra do indeed produce smaller deposition values. This is due to the fact that a harder spectrum produces more ionization at lower altitudes, which means that the $NO_y$ compounds are more likely to be incorporated into precipitation rather than being photolyzed, as they are at higher altitudes.

5.5 Enhancement of Ground-Level Solar UV

Finally, it is interesting to consider the effect on ground-level ultraviolet radiation under the ozone depletion modeled here. First, the globally averaged depletion found here is very similar to the current anthropogenic depletion, though the local maximum depletion in this case is much smaller than that found under the ozone "hole" over Antarctica. A quantity of interest to human populations when discussing ground-level UV is the "UV Index" [WMO 1994]. This index is intended to be a simple way to communicate with the public the danger posed by UV radiation at any given time and place. It is based primarily on a weighting function for damage to human skin (sunburn). We use the UV index here, then, as a simple way of quantifying the potential impact on human populations due to the ozone depletion in our results.

In order to compute the UV index at the surface under our modeled atmospheric conditions, we use version 4.6 of the publically available Tropospheric Ultraviolet and Visible (TUV) atmospheric radiative transfer model, downloaded from http://cprm.acd.ucar.edu/Models/TUV/ [*Madronich and Flocke,* 1997]. In order to make TUV easier to use with output from the GSFC model we modified it slightly to read in an altitude profile of $O_3$ number density from the ground to 90 km (with 46 levels). We then used a script to automatically run TUV at local noon for every daily output point at each of the 18 latitude points used in the GSFC model. After running TUV in this manner for both an SPE case and a background case, we computed the percent increase in UV index values at each time and latitude point.

The results for the September 1989 case are shown in Figure 14. In this figure we restrict our attention to latitudes between 35° and 65° North, since the $O_3$ change is greatest in the North in our results, and most human populations which would be affected are located in this region. An increase of 10-20% is evident across much of this region for more than a year, with the highest increase at the most northerly latitudes. For reference, the maximum UV index at 35° North in the unperturbed model run (without SPE ionization) is about 9.0; this corresponds to the "very high" danger category. A 10-20% increase could move the index from "very high" to "extreme" at some points, though this is not a particularly dramatic change.

6. Summary and Conclusions

We have completed modeling of the atmospheric changes due to the 1859 Carrington solar proton event, using four recent events as proxy cases. Results of our modeling, particularly changes in column $O_3$ values, are consistent with past results reported in *Thomas et al.* [2007] and *Rodger et al.* [2008]. However, differences do appear, due to the different modeling techniques used here as compared to those previous studies. Notably, we find a greater spread in the maximum $O_3$ column density change for the different proxy spectra than that found in *Rodger et al.* [2008]. Our results indicate that the details of the proton spectrum matter, though the differences explored here do not result in wildly different results. In particular it may be noted that a harder proton spectrum has a more significant impact on stratospheric $O_3$, leading to greater overall column density changes. Nitrate deposition computed here shows a similar difference, as may be expected, and values compare well to those previously found. We have also explored the increase in potentially harmful UV radiation at the ground under the depleted $O_3$ conditions. We find it to be increased, but not significantly enough to cause major changes in human health (nor would we expect an observable impact on other surface-dwelling life).

Overall, we have shown that long-term global atmospheric changes caused by SPEs should ideally be studied using models that include the full altitude range (especially altitudes within and below the stratosphere), as well as a spatial extent that covers at least the full latitude range. Changes will of course depend on longitude as well, but on time scales longer than a few weeks those changes are negligible compared to latitude differences. Additionally, we find that the details of the proton spectrum do have an impact on the atmospheric effects, though the differences considered here are not extreme. It should be noted that only one possible proton spectrum form was used here (the Weibull distribution). Other forms are possible. In fact, the Band spectrum (which consists of two smoothly connected power laws [*Band et al.*, 1993]) may be more realistic [*Tylka and Dietrich,* 2009]. Future studies should take this into account and examine the impact of different spectral distributions as well as changes in specific parameter values. With the increasing likelihood of major solar events like the Carrington accurate modeling of the effects considered here, as well as others, becomes even more important.

Acknowledgements

This work has been supported by the National Aeronautics and Space Administration under grant No. NNX09AM85G, through the Astrobiology: Exobiology and Evolutionary Biology Program. Computational time for this work was provided by the High Performance Computing Environment (HiPACE) at Washburn University; thanks to Dr. Steve Black for invaluable assistance with computing resources. Thanks to Craig Rodger for providing a data table with the intensity-time profile used here. Thanks also to Pekka Verronen for data tables of atmospheric ionization that were helpful in validating our computations. The authors are grateful to Sasha Madronich for assistance with using and modifying the TUV code.

Figures

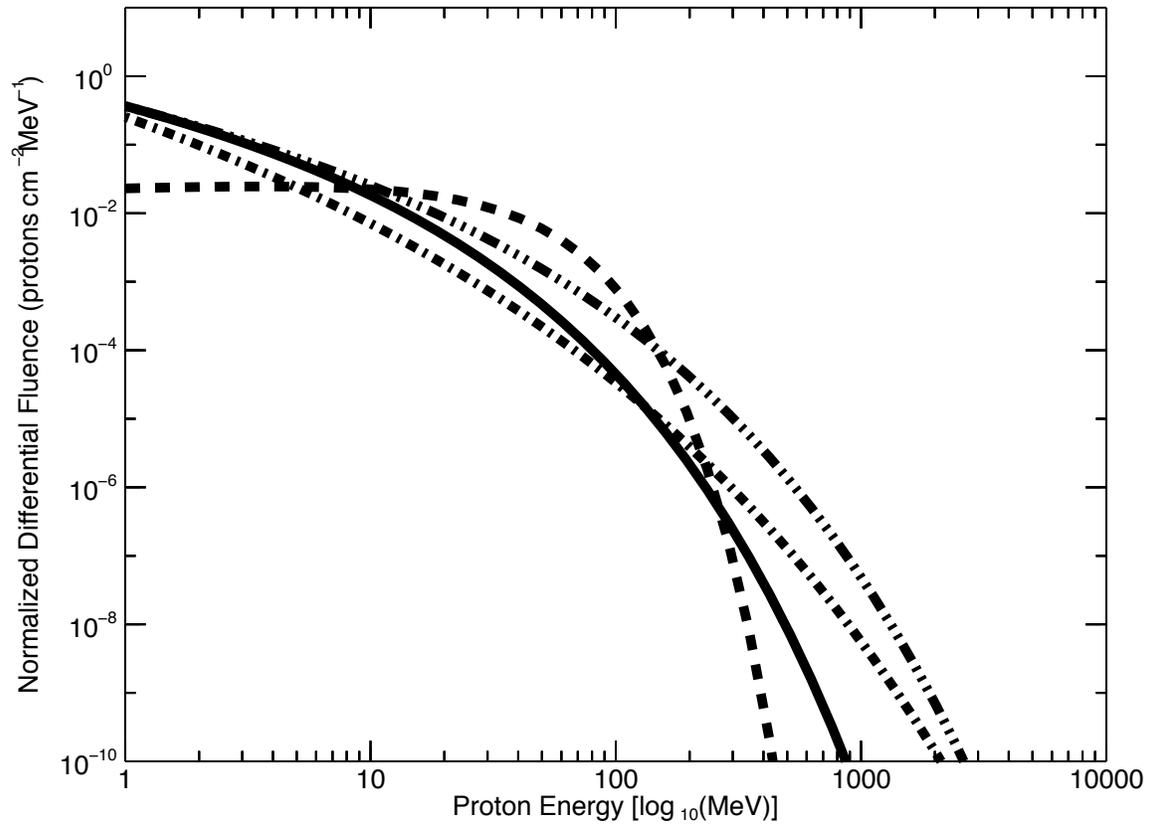

**Figure 1 – Normalized differential fluence functions for the four SPE cases. Solid line = March 1991, Dashed line = August 1972, Dash dot line = October 1989, Dash three-dot line = September 1989. Each curve is normalized to a total *E* > 30 MeV proton fluence of 1.9 × 10$^{10}$ cm$^{-2}$.**

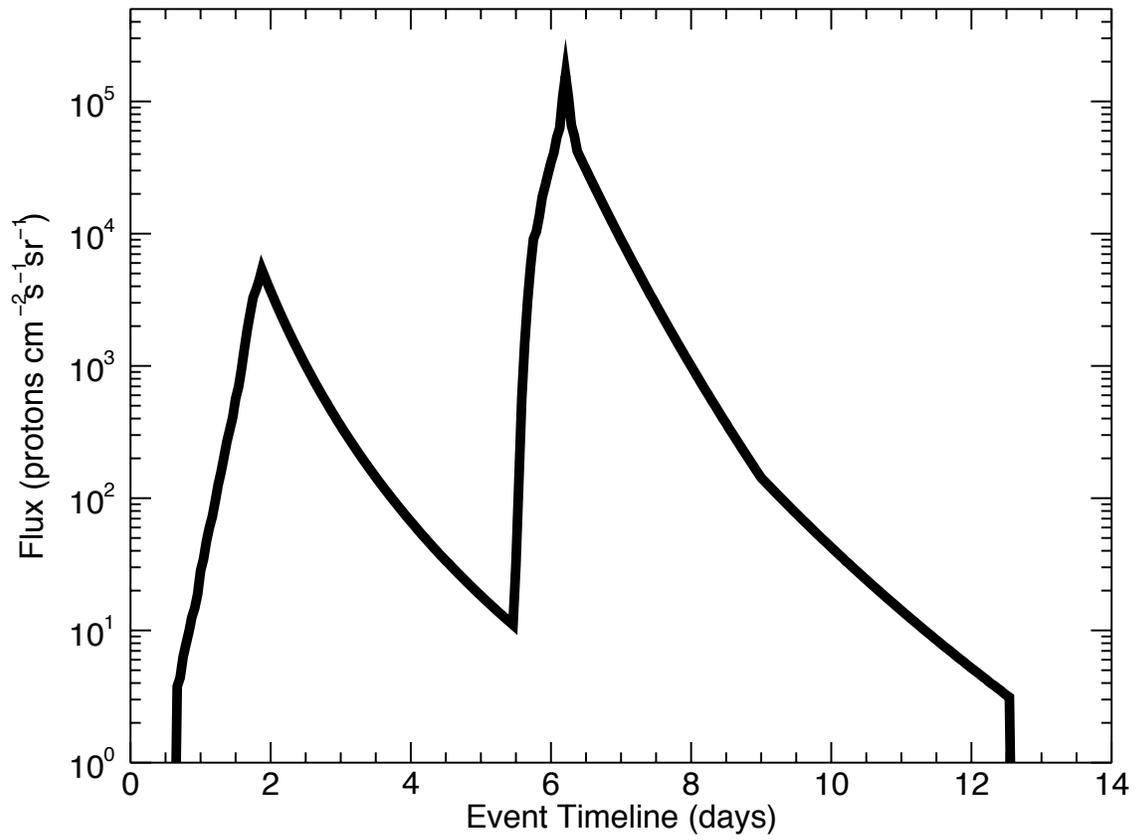

**Figure 2 – Intensity-time profile of proton flux, following *Smart et al.* [2006]. Day 1 here corresponds to 28 August, 1859.**

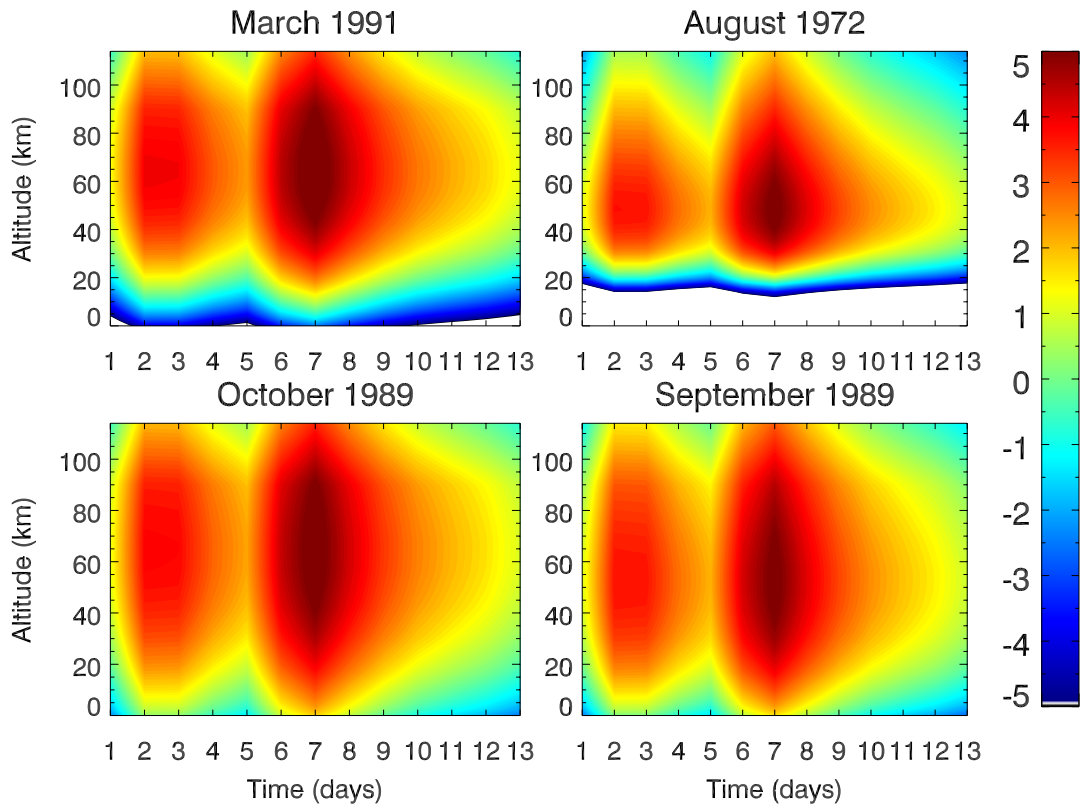

**Figure 3 – Ionization rate (ions cm$^{-3}$ s$^{-1}$) as a function of altitude and time for the four SPE cases. Day 1 here corresponds to 28 August 1859.**

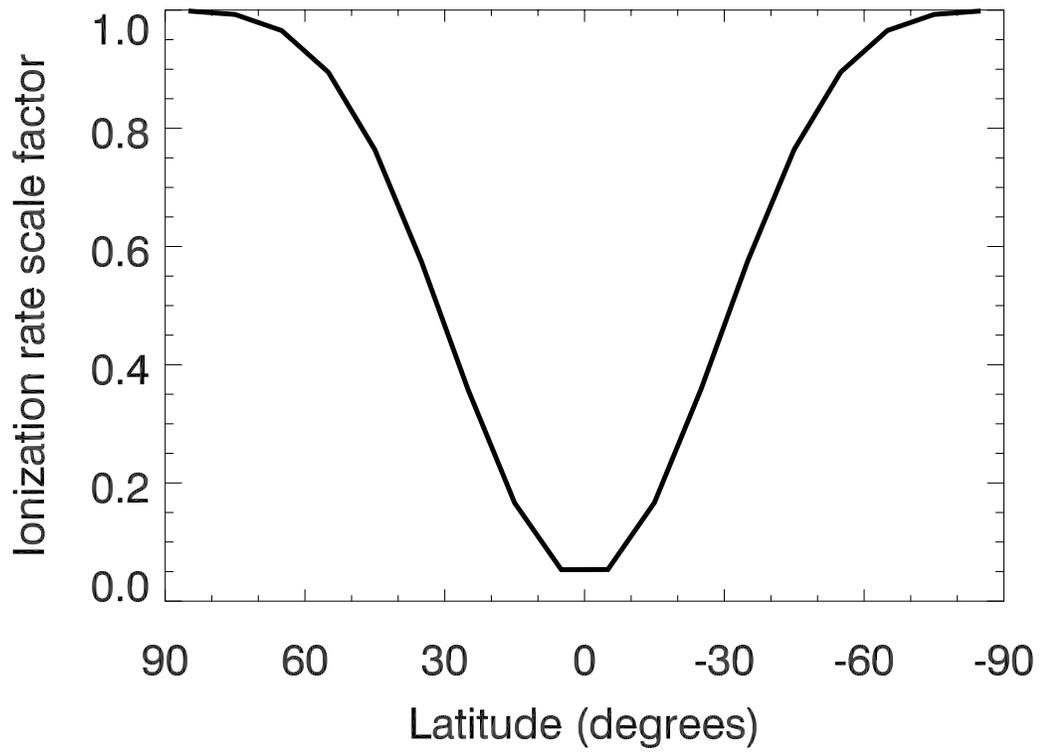

**Figure 4 – Scale factor for ionization as a function of latitude.**

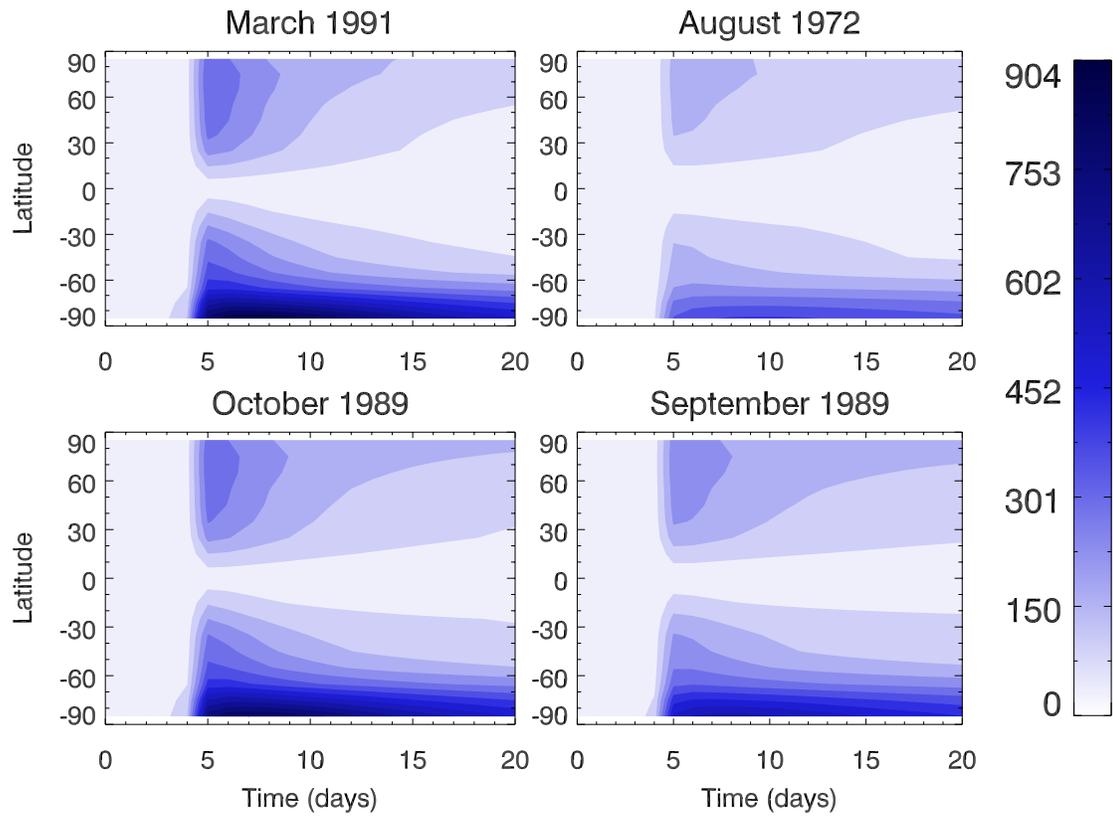

**Figure 5** – Percent difference in column NO$_y$ between perturbed and unperturbed runs for each proxy case over the first 20 days. Time zero corresponds to 27 August 1859.

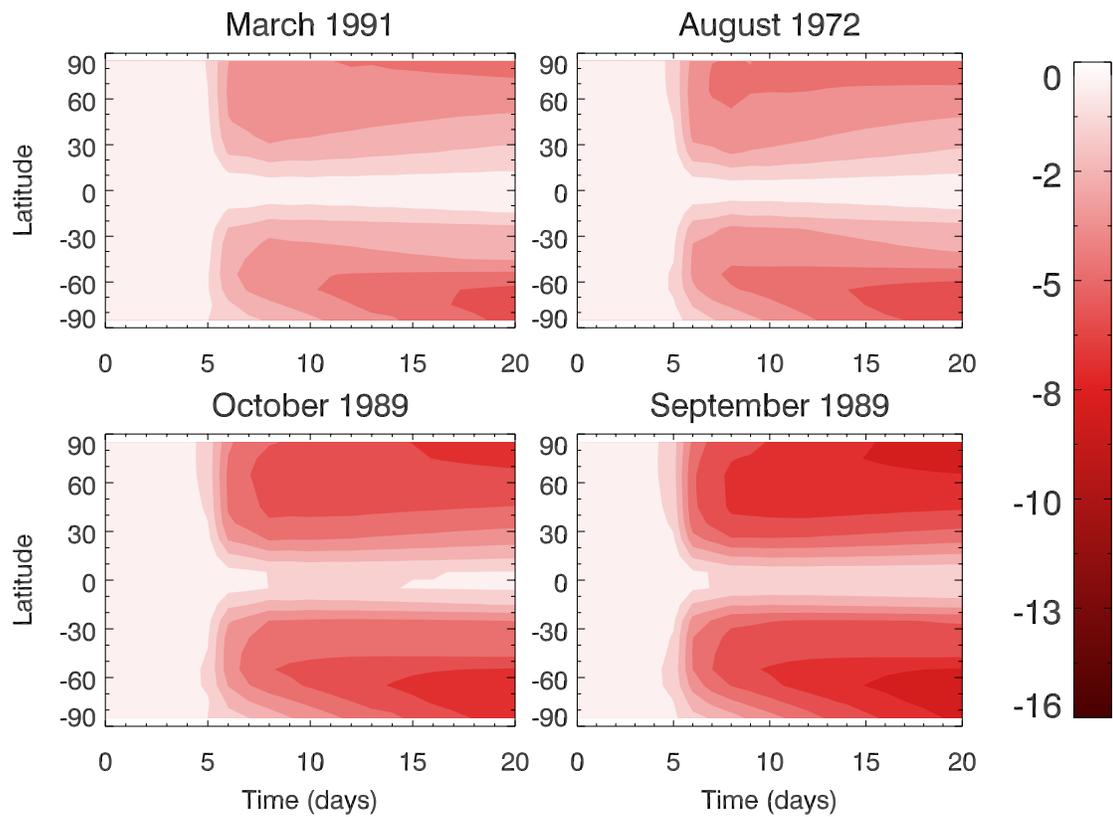

**Figure 6** - Percent difference in column O$_3$ between perturbed and unperturbed runs for each proxy case over the first 20 days. Time zero corresponds to 27 August 1859.

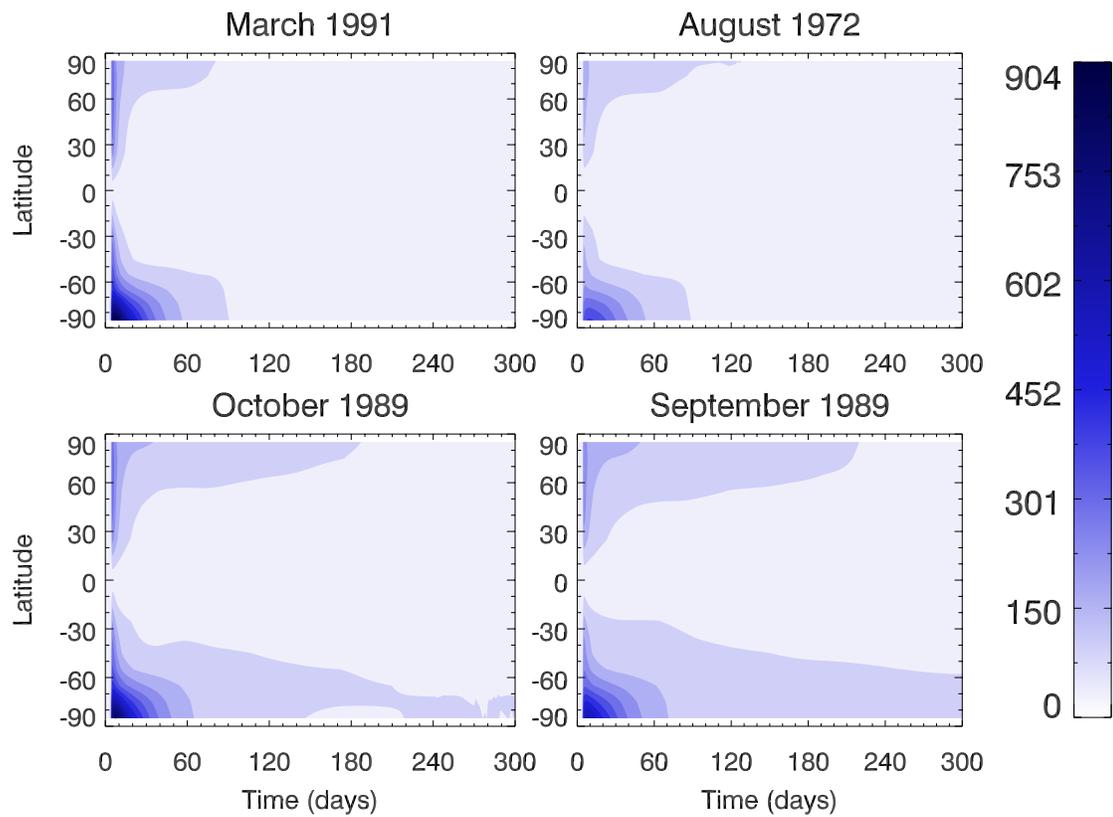

**Figure 7 – Similar to Figure 5, percent difference in column $NO_y$ between perturbed and unperturbed runs for each proxy case over the first 300 days.**

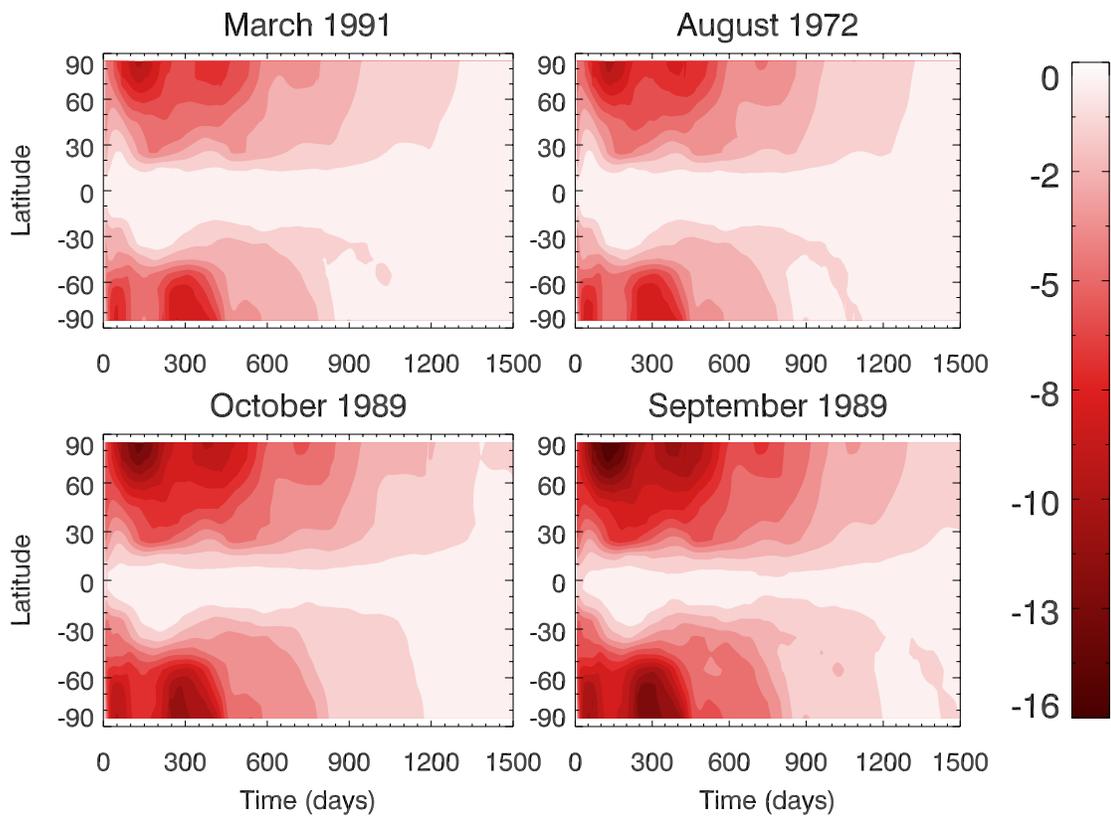

**Figure 8 - Similar to Figure 6, percent difference in column O$_3$ between perturbed and unperturbed runs for each proxy case over the first 1500 days.**

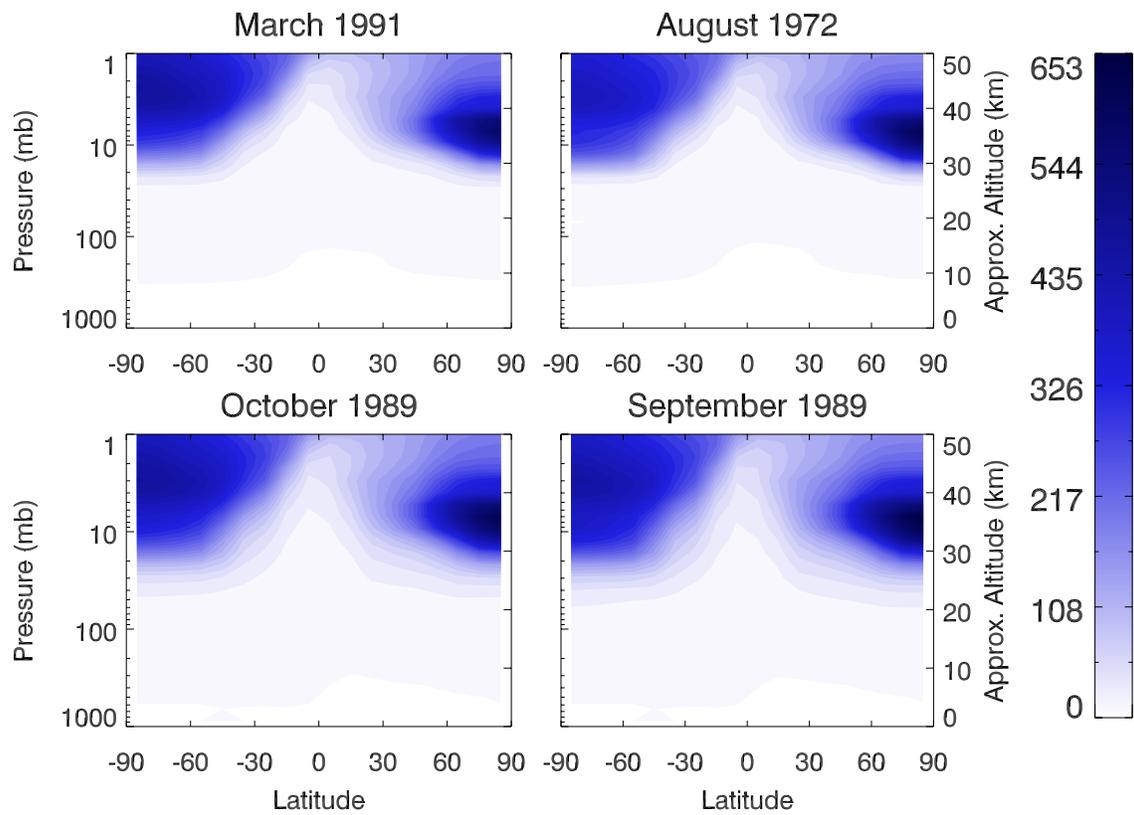

**Figure 9** – Percent difference in volume number density of $NO_y$ as a function of altitude and latitude at 100 days after the start of the event. The altitude range is from the ground to the top of the stratosphere.

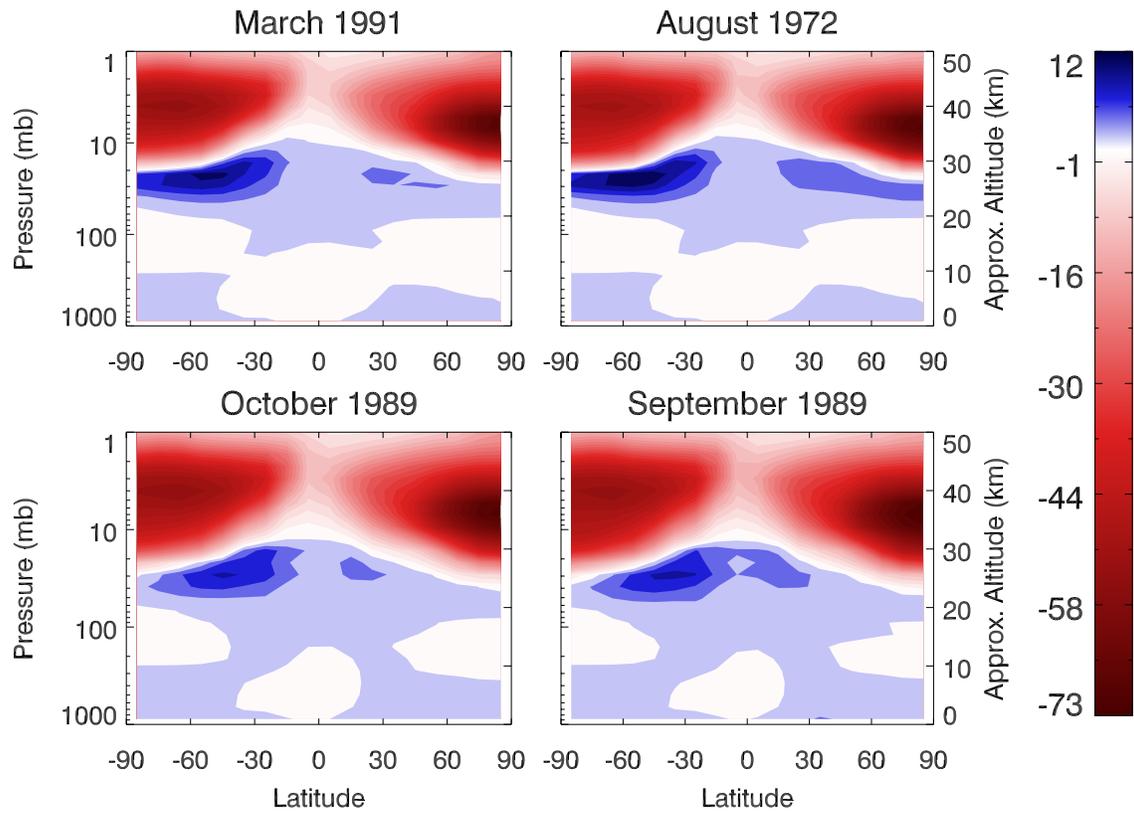

**Figure 10 - Percent difference in volume number density of $O_3$ as a function of altitude and latitude at 100 days after the start of the event. The altitude range is from the ground to the top of the stratosphere.**

Revisiting the Carrington Event 29

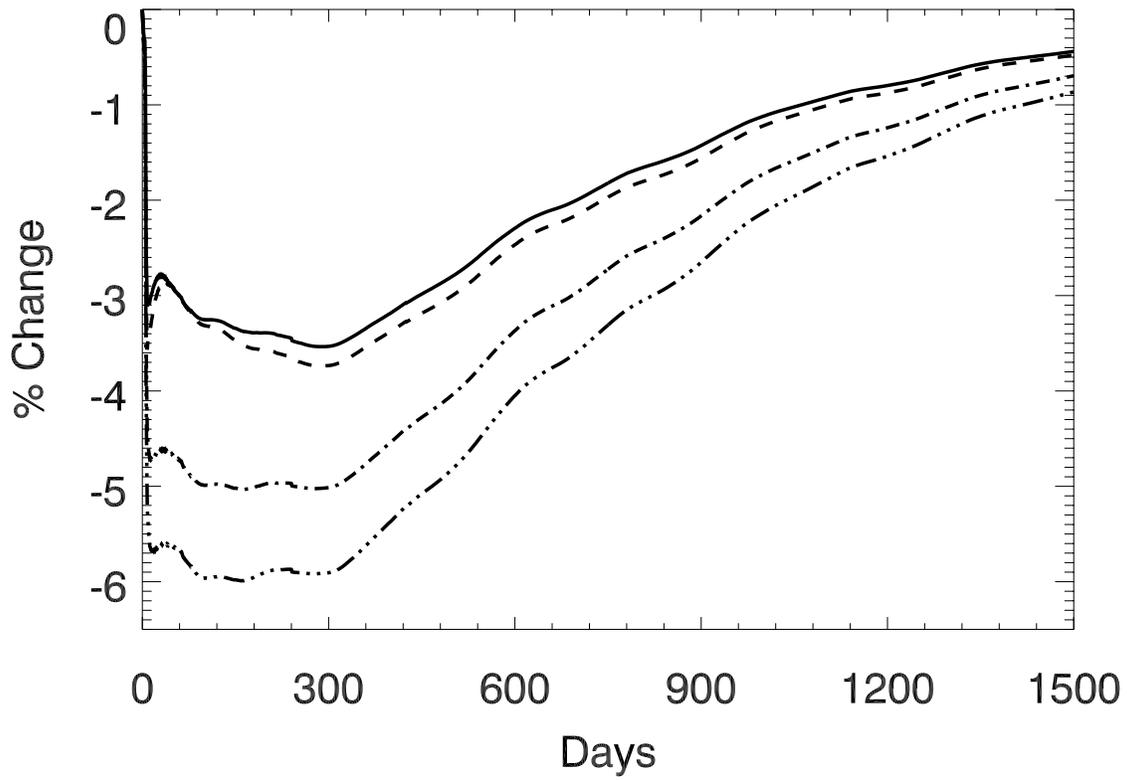

**Figure 11 – Globally averaged percent difference in $O_3$ column density between perturbed and unperturbed runs for each proxy case over the first 1500 days. Solid line = March 1991, Dashed line = August 1972, Dash dot line = October 1989, Dash three-dot line = September 1989.**

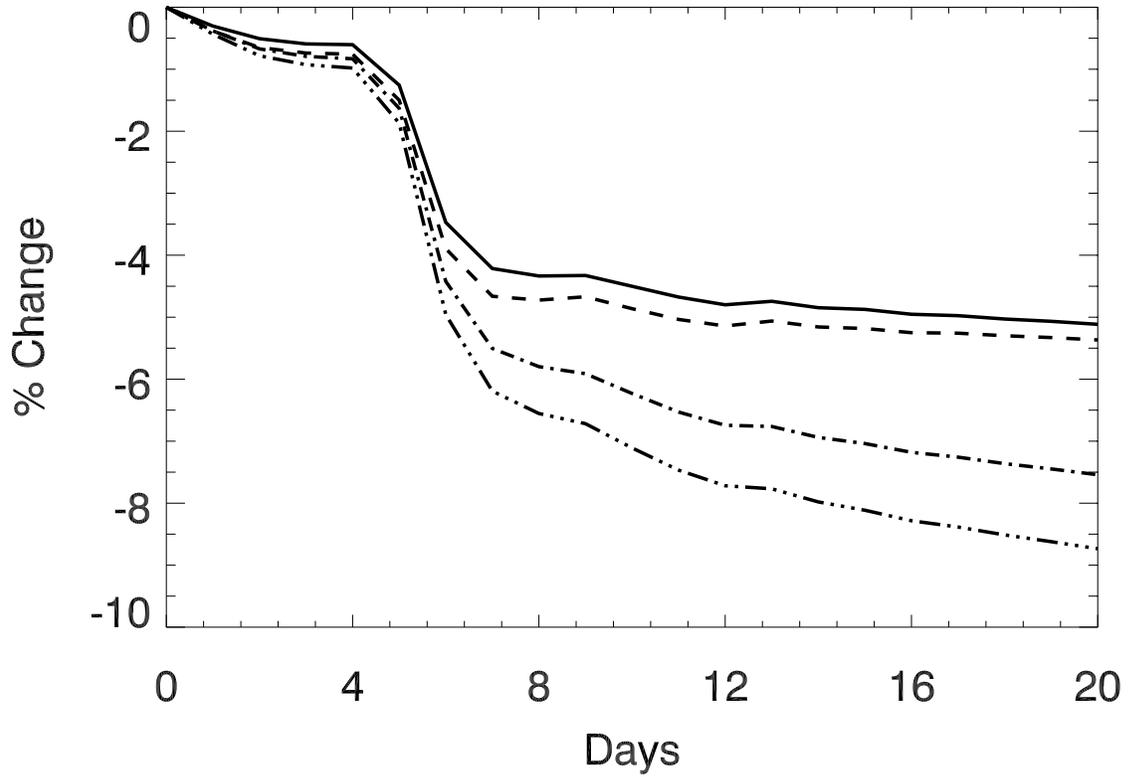

**Figure 12 - Percent difference in total O$_3$ column density at 75° North for the first 20 days. Solid line = March 1991, Dashed line = August 1972, Dash dot line = October 1989, Dash three-dot line = September 1989.**

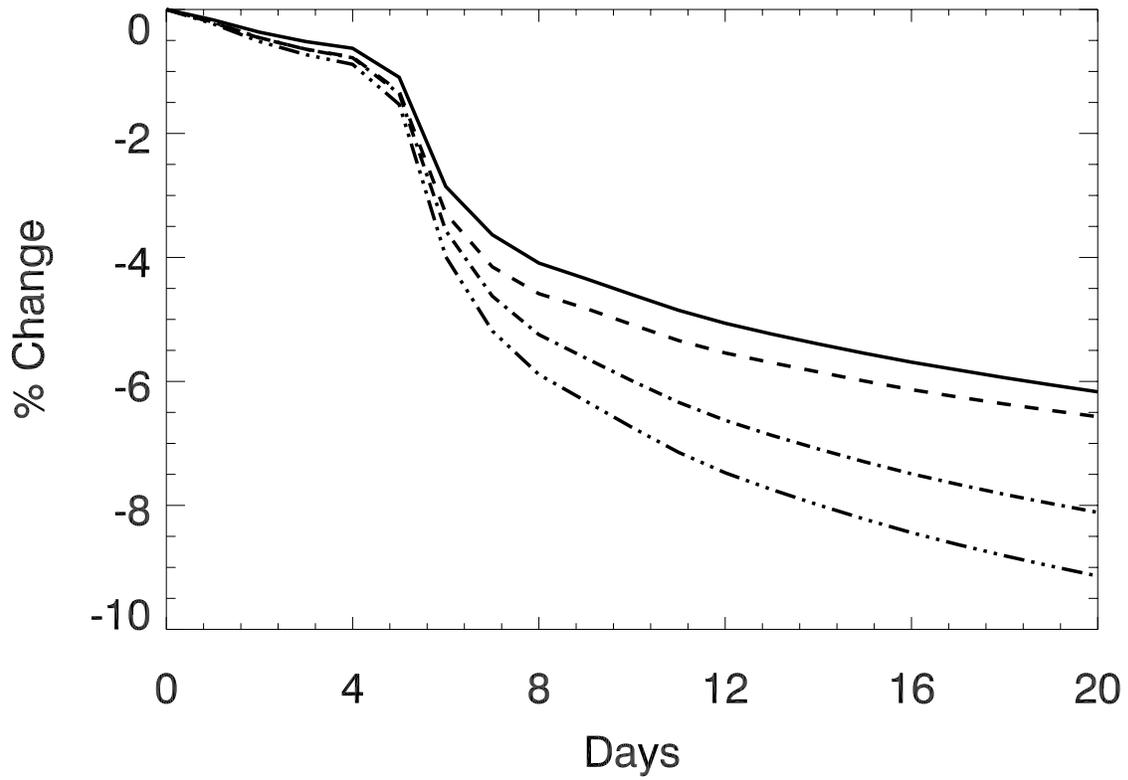

**Figure 13** – Same as Figure 12, but for 75° South. Solid line = March 1991, Dashed line = August 1972, Dash dot line = October 1989, Dash three-dot line = September 1989.

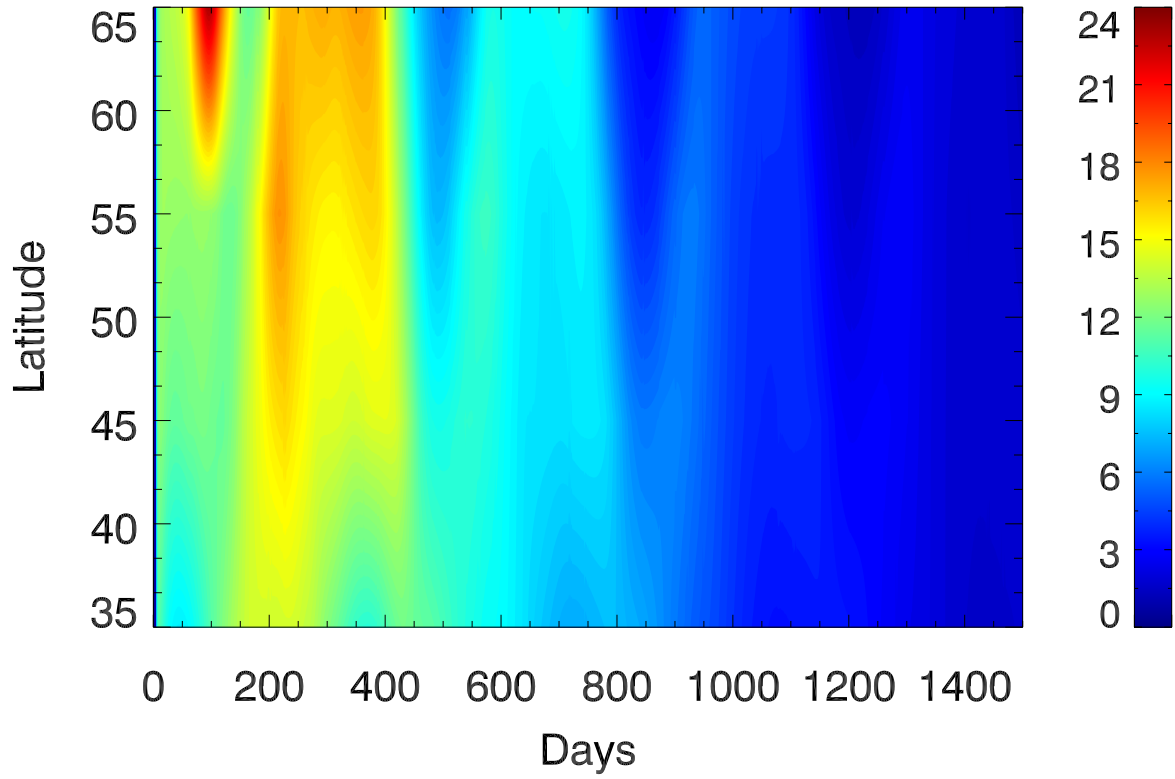

**Figure 14 - Percent increase in UV index values for the September 1989 case.**

Tables

Table 1. Weibull Fitting Parameters

| SPE Date | $k$ | $\alpha$ |
|---|---|---|
| 4 August 1972 | 0.0236 | 1.108 |
| 29 September 1989 | 0.877 | 0.3841 |
| 19 October 1989 | 2.115 | 0.2815 |
| 23 March 1991 | 0.972 | 0.441 |

Table 2. Maximum ionization rate, $NO_y$ column density percent change, $O_3$ column density percent change and global average $O_3$ column density percent change for each modeled case.

| Proxy Case | Max ionization rate (ions cm$^{-3}$ s$^{-1}$ per day) | Max $NO_y$ column % change | Max $O_3$ column % change | Max global average $O_3$ column % change |
|---|---|---|---|---|
| March 1991 | $2.46 \times 10^5$ | 905 | -10.8 | -3.54 |
| August 1972 | $1.46 \times 10^5$ | 398 | -11.0 | -3.74 |
| October 1989 | $1.94 \times 10^5$ | 840 | -14.5 | -5.04 |
| September 1989 | $1.43 \times 10^5$ | 606 | -16.4 | -5.99 |